\documentclass[nofootinbib,a4paper,aps,pra,reprint,twocolumn,preprintnumbers,amsmath,amssymb,10pt, superscriptaddress]{revtex4-2}

\usepackage{amsmath}
\usepackage{verbatim}
\usepackage{graphicx}
\usepackage{color}
\usepackage{tikz}
\usepackage{hyperref}
\usepackage{braket, enumitem,}
\usepackage[normalem]{ulem}
\usepackage{upgreek}

\begin{document}

\title{Gapless deconfined phase in a $\mathbb{Z}_N$ symmetric Hamiltonian created in a cold-atom setup}

\author{Mykhailo V. Rakov}
\affiliation{Instytut Fizyki Teoretycznej, Uniwersytet Jagiello\'nski, ulica Łojasiewicza 11, 30-348 Krak\'ow, Poland}

\author{Luca Tagliacozzo}
\affiliation{Instituto de Fisica Fundamental IFF-CSIC, Calle Serrano 113b, Madrid 28006, Spain}

\author{Maciej Lewenstein}
\affiliation{Institut de Ciencies Fotoniques (ICFO), Av. Karl Friedrich Gauss 3, 08860 Barcelona, Spain}
\affiliation{ICREA, Pg. Lluís Companys 23, 08010 Barcelona, Spain}

\author{Jakub Zakrzewski}
\affiliation{Instytut Fizyki Teoretycznej, Uniwersytet Jagiello\'nski, ulica Łojasiewicza 11, 30-348 Krak\'ow, Poland}
\affiliation{Mark Kac Complex Systems Research Center, Uniwersytet Jagiello{\'n}ski,30-348  Krak{\'o}w, Poland}

\author{Titas Chanda}
\affiliation{Department of Physics, Indian Institute of Technology Indore, Khandwa Road, Simrol, Indore 453552, India}
\affiliation{Department of Physics, Indian Institute of Technology Madras, Chennai 600036, India}

\begin{abstract}
We investigate a quasi-two-dimensional system consisting of two species of alkali atoms confined in a specific optical lattice potential [Phys. Rev. A \textbf{95}, 053608 (2017)]. In the low-energy regime, this system is governed by a unique $\mathbb{Z}_N$ gauge theory, where field theory arguments have suggested that it may exhibit two exotic gapless deconfined phases, namely a dipolar liquid phase and a Bose liquid phase, along with two gapped (confined and deconfined) phases. We address these predictions numerically by using large-scale density matrix renormalization group simulations. Our findings provide conclusive evidence for the existence of a gapless Bose liquid phase for $N \geq 7$. We demonstrate that this gapless phase shares the same critical properties as one-dimensional critical phases, resembling weakly coupled chains of Luttinger liquids. In the range of ladder and cylinder geometries and $N$ considered, the gapless dipolar phase predicted theoretically is still elusive and its characterization will probably require a full two-dimensional treatment.
\end{abstract}

\maketitle

\section{Introduction}

Gauge-invariant theories form the bedrock of our current understanding of the fundamental forces of nature, describing three of the four known forces: electromagnetic, weak, and strong interactions~\cite{peskin_book, Schwartz2013, Tanabashi2018_2, tong}. They are also crucial for the low-energy description of topological order in certain materials within condensed matter physics~\cite{Wen2007, Altland2010, Fradkin2013}.
While their Lagrangian formulations on lattices~\cite{Wilson1974, Susskind1979, Kogut1983, Svetitsky1986, McLerran1986, DeTar2009, Fodor2012}  are well-established through symmetry principles, our understanding of their phase diagrams remains incomplete, particularly in regimes of strong correlation where traditional Monte Carlo lattice methods (for a U.S.-centric perspective on the topic, see~\cite{Davoudi2022}) are ineffective. This limitation poses a significant hindrance in the study of nuclear matter at high fermionic densities, which is crucial for understanding of the interiors of neutron stars. Besides finite fermionic densities, other significant challenges include real-time dynamics which are necessary for understanding the scattering processes in heavy-ion collisions, and topological terms, which could provide insights into the asymmetry between matter and antimatter in the universe.

In light of these challenges, the quantum simulation community, following Feynman's intuition~\cite{Feynman1982}, has begun exploring methods to simulate gauge theories using ultracold atoms, trapped ions, and superconducting qubits \cite{Zohar2012, Tagliacozzo2013, Tagliacozzo2013a, Wiese2013, Zohar2015, Dalmonte2016, Banuls2020, Aidelsburger2021}. Recent advancements have refined many of these initial proposals, and the first experimental realizations of one-dimensional gauge systems have been achieved~\cite{Martinez2016, Schweizer2019, Yang2020, Nguyen2022, Riechert2022, Chisholm2022, Bauer2023}.
These experiments include demonstrations of local gauge invariance in large Bose-Hubbard-like systems.
Most of these proposals rely on the microscopic realization of gauge theories, enforcing local gauge invariance and protecting it from environmental effects. However, such approaches face significant challenges, as ultracold atoms are typically dilute and neutral, and their interactions are limited to two-body interactions. Consequently, implementing the higher-body interactions, necessary for gauge Hamiltonians in two or higher dimensions, is difficult. Even though such higher-order interactions are hard to obtain, advanced experimental platforms, such as Rydberg atoms, where tweezers can manipulate atoms sequentially to achieve higher-order interactions through a series of two-body interactions in the presence of the so-called Rydberg-blockade, can provide access to them ~\cite{Tagliacozzo2013, Tagliacozzo2013a,Surace2020, Cheng2023, Homeier2023, Xu2024, Meurice2024}. 

A different approach for simulating  gauge theories is based  on the observation that many condensed matter systems can be described by gauge theories at low energies. This emergent realization of gauge theories contrasts with the microscopic implementations discussed above. Examples of such systems include topological spin liquids in highly frustrated magnets and quantum Hall states \cite{broholm2020}.

This alternative approach was first explored in Ref.~\cite{Dutta2017}, where the authors proposed an experimental protocol for realizing emergent Abelian lattice gauge theories using a mixture of two species of ultracold alkali atoms trapped on an optical lattice. A generalized Bose-Hubbard Hamiltonian can be created by fast modulation of interspecies interactions,  giving rise to exotic, anisotropic gauge theories at low energies. In two dimensions (2D), such exotic gauge theories have phase diagrams that differ significantly from their one-dimensional counterparts.

While Ref.~\cite{Dutta2017} presented the experimental protocol and a qualitative phase diagram, our work revisits the same system, providing quantitative results on some phases using tensor network (TN)~\cite{Schollwoeck2011, Orus2014, Ran2020} simulations. 
In recent years, TN methods, free from the limitations associated with traditional Monte Carlo approaches, have excelled in low-dimensional lattice gauge theory calculations~\cite{banuls_jhep_2013, buyens_prl_2014, kuhn_pra_2014, banuls_prd_2015, buyens_prd_2016, pichler_prx_2016, banuls_prl_2017, buyens_prx_2016, chanda20, Magnifico2020, Felser2020, chanda2021, Montangero2021, Magnifico2021, chanda2024, Cataldi2024, Magnifico2024}.
In this work, we consider infinite cylinders with different finite radii, characterizing the liquid phases with emergent discrete $\mathbb{Z}_N$ symmetry and identifying phase boundaries between topological and trivial phases for various $N$ and cylinder lengths, corroborating the predictions from Ref.~\cite{Dutta2017}.
Specifically, we demonstrate that for $\mathbb{Z}_N$ emergent gauge theories with $N \leq 6$, the system exhibits two gapped phases: the confined and deconfined phases, with the latter possessing non-trivial topological order. However, for $N \geq 7$, a gapless liquid (deconfined) phase emerges between these gapped phases. Interestingly, we find that this gapless phase exhibits critical properties akin to one-dimensional (1D) Bose liquids, characterized by weakly coupled chains of identical Luttinger liquids.

The paper is organized as follows. In Sec.~\ref{sec:model} we introduce the two-species cold-atomic setup and describe the emergent $\mathbb{Z}_N$ gauge theories due to inter-species dynamics in the low-energy description. We also present the global $\mathbb{Z}_N$-symmetric model in the dual lattice. In Sec.~\ref{sec:phase}, we examine the phase diagram of the system for different quasi-2D ladder and cylinder geometries through the lens of entanglement entropy and correlation length. In Sec.~\ref{sec:characterize}, we analyze the gapless deconfined phase by examining different correlation lengths and the scaling of entanglement entropy. Finally, we draw our conclusion in Sec.~\ref{sec:conclu}.

\begin{figure*}[htb]
    \includegraphics[width = \linewidth]{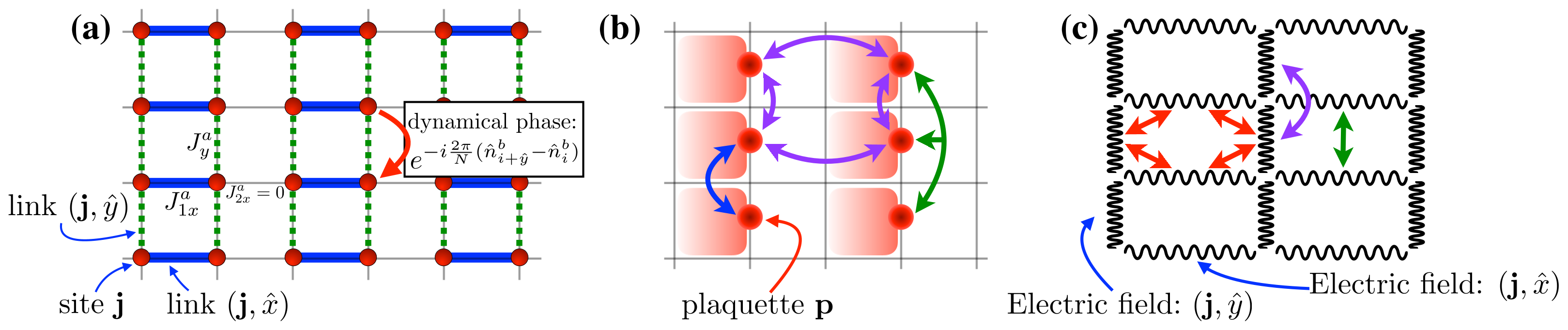}
\caption{(Color online.) The schematic depiction of the system considered in this work. (a) The two-species Bose-Hubbard system in a square superlattice described by Eq.~\eqref{eq:BH}. The nearest-neighbor tunneling of the $a$-bosons along the $y$-direction is modulated by the occupation of the $b$-bosons on the neighboring sites. (b) The effective Hamiltonian (Eq.~\eqref{eq:Hamil_dual} or \eqref{eq:Hamil_final}) on the coarse-grained lattice where two plaquettes along the $x$ direction are combined into one. Here the relevant degrees of freedom sit on the centers of the $2 \times 1$ square as depicted by the red circles. The blue, green, and purple arrows depict the two-, three-, and four-plaquette interactions as  described by the second, third, and fourth terms respectively in Eq.~\eqref{eq:Hamil_dual} or \eqref{eq:Hamil_final}.
(c) The effective $\mathbb{Z}_N$ gauge theory described by Eq.~\eqref{eq:Hamil_gauge} or \eqref{eq:Hamil_gauge_final} obtained via duality transformation as mentioned in the text. The wiggly lines describe the gauge or electric-field degrees of freedom located on the bonds of the coarse-grained lattice. The red arrows denote the magnetic plaquette term (the first term in Eq.~\eqref{eq:Hamil_gauge} or \eqref{eq:Hamil_gauge_final}), while the green and purple arrows describe the two-body interaction along the $y$-direction given by the third and fourth terms respectively in the Hamiltonian \eqref{eq:Hamil_gauge} or \eqref{eq:Hamil_gauge_final}.
}
\label{fig:schem}
\end{figure*}

\section{The System}
\label{sec:model}

Following the approach presented in Ref.~\cite{Dutta2017}, we consider emergent Abelian lattice gauge theories within the low-energy manifold of two-species cold-atom dynamics in 2D. The setup involves two bosonic species, referred to as $a$ and $b$ bosons, on a 2D optical superlattice. A specific configuration of link-dependent tunneling for the $a$-bosons  and the fast modulation of the interactions between the two species produces the desired effect. Namely, the $b$-bosons behave as gauge bosons at low energies. The Hamiltonian  describing the combined behavior  of the $a$ and $b$ bosons is given by a generalization of the Bose-Hubbard model:
\begin{align}
\hat{H}_{BH} = & - J^b \sum_{\mathbf{j}, \hat{\delta}} \left( \hat{b}^{\dagger}_{\mathbf{j}}\hat{b}_{\mathbf{j} + \hat{\delta}} + \text{h.c.} \right) 
- J_{1x}^a  \sum_{j_x \in \text{odd}} \left(\hat{a}^{\dagger}_{\mathbf{j}}\hat{a}_{\mathbf{j} + \hat{x}} + \text{h.c.}\right) \nonumber \\
&- J^a_y\sum_{j_y} \left(\hat{a}^{\dagger}_{\mathbf{j}} e^{i\frac{2\pi}{N}(\hat{n}^b_{\mathbf{j} + \hat{y}} - \hat{n}^b_{\mathbf{j}})} \hat{a}_{\mathbf{j} + \hat{y}} + \text{h.c.}\right) \nonumber \\
& + \frac{U}{2}\sum_{\mathbf{j}} (\hat{a}^{\dagger}_{\mathbf{j}})^2 \hat{a}_{\mathbf{j}},
\label{eq:BH}
\end{align}
where $N$ is a positive integer. For the schematic depiction of the system, see Fig.~\ref{fig:schem}(a). The first term in the Hamiltonian describes the nearest-neighbor tunneling of the $b$-bosons, which is uniform in both the $x$ and $y$ directions with an amplitude $J^b$ (here $\hat{\delta} = \hat{x}$ or $\hat{y}$). The second term represents the nearest-neighbor tunneling of the $a$-bosons along the $x$-direction with an amplitude $J^a_{1x}$ between sites $\mathbf{j}$ and $\mathbf{j} + \hat{x}$ for odd $j_x$. Here, the tunneling amplitude of $a$-bosons between sites $\mathbf{j}$ and $\mathbf{j} + \hat{x}$ for even $j_x$ is considered vanishingly small compared to the other terms in the Hamiltonian, and that is why it is neglected. The third term describes the tunneling of $a$-bosons in the $y$-direction with an amplitude $J^a_y$, which is modulated by the occupation of $b$-bosons on the neighboring sites (see Fig.~\ref{fig:schem}(a)). The final term denotes the onsite repulsion between $a$-bosons. Here, we consider the limit $U \gg J^a_{1x}, J^a_y$, so that at half-filling, the $a$-bosons effectively behave as hardcore bosons. Moreover, the onsite repulsion between $b$-bosons is considered to be vanishingly small, so we have a large occupation of the $b$-bosons (i.e., $\bar{n}_b \gg 1$) at each site.

In the limit $J^a_y \ll J^a_{1x}$, the dynamics of $a$-bosons creates effective plaquette interactions for the $b$-bosons in a superlattice where plaquette $\mathbf{p}$ (red shadowed region in Fig.~\ref{fig:schem}(b)) contains the sites $\mathbf{j}$, $\mathbf{j} + \hat{x}$, $\mathbf{j} + \hat{x} +\hat{y}$, $\mathbf{j} + \hat{y}$ for odd $j_x$ (i.e., every second plaquette of the original lattice, for details, see \cite{Dutta2017}). We define a plaquette operator $\hat{\mathcal{B}}_{\mathbf{p}}$ as
\begin{equation}
    \hat{\mathcal{B}}_{\mathbf{p}} = \frac{2\pi}{N}  \left( \hat{n}^b_{\mathbf{j}} - \hat{n}^b_{\mathbf{j}+\hat{x}} +  \hat{n}^b_{\mathbf{j}+\hat{x} +\hat{y}} - \hat{n}^b_{\mathbf{j}+\hat{y}} \right),
\end{equation}
with $\hat{n}^b_{\mathbf{j}}$ being the number operator for $b$-bosons, so that $e^{i\hat{\mathcal{B}}_{\mathbf{p}}}$ operators have eigenvalues $\left\{\exp\left(i\frac{2\pi n}{N}\right)\right\}$ with $n \in \left[-\frac{(N-1)}{2}, -\frac{(N-2)}{2}, \dots, \frac{(N-2)}{2}, \frac{(N-1)}{2} \right]$. In the second order perturbation theory, the effective Hamiltonian of the $b$-bosons dressed by the dynamics of $a$-bosons reads as (Fig.~\ref{fig:schem}(b)):
\begin{align}
    \hat{H}_P &= - 2 \sum_{\mathbf{p}} \cos\hat{\mathcal{B}}_{\mathbf{p}} 
    -2 g^2 \sum_{\mathbf{p}} \Bigg[ \cos\left(\frac{4\pi}{N} \left(\hat{\mathcal{L}}_{\mathbf{p}}-\hat{\mathcal{L}}_{\mathbf{p}-\hat{y}}\right) \right) \nonumber \\
    &+2\cos\left(\frac{2\pi}{N}  \left(2\hat{\mathcal{L}}_{\mathbf{p}}-\hat{\mathcal{L}}_{\mathbf{p}+\hat{y}}-\hat{\mathcal{L}}_{\mathbf{p}-\hat{y}}\right)\right)    \nonumber \\
    &+\cos\left(\frac{2\pi}{N} \left(\hat{\mathcal{L}}_{\mathbf{p}}-\hat{\mathcal{L}}_{\mathbf{p}+\hat{x}}-\hat{\mathcal{L}}_{\mathbf{p}+\hat{y}}+\hat{\mathcal{L}}_{\mathbf{p}+\hat{x}+\hat{y}}\right)\right) \Bigg],
    \label{eq:Hamil_dual}
\end{align}
where the coupling constant $g$ depends on $b$-boson filling as well as tunneling amplitudes of both kinds of bosons as $g^2=\frac{\bar{n}_b}{2} \frac{J^{a}_{1x}}{J^b}$. Here the plaquette operator $\hat{\mathcal{L}}_{\mathbf{p}}$ satisfies the following algebra:
\begin{align}
\left[\hat{\mathcal{L}}_{\mathbf{p}}, e^{\mp i \hat{\mathcal{B}}_{\mathbf{p}}} \right] &= \pm e^{\mp i \hat{\mathcal{B}}_{\mathbf{p}}}, \nonumber \\
\left[\hat{\mathcal{B}}_{\mathbf{p}}, e^{\pm i \frac{2\pi}{N}\hat{\mathcal{L}}_{\mathbf{p}}} \right] &= \pm \frac{2\pi}{N} e^{\pm i \frac{2\pi}{N}\hat{\mathcal{L}}_{\mathbf{p}}},
\end{align}
i.e., $e^{\pm i \frac{2\pi}{N}\hat{\mathcal{L}}_{\mathbf{p}}}$ fulfills the role of $\mathbb{Z}_N$ ladder operators for the eigenstates of $\hat{\mathcal{B}}_{\mathbf{p}}$, and conversely $e^{\mp i \hat{\mathcal{B}}_{\mathbf{p}}}$ for the eigenstates of  $\hat{\mathcal{L}}_{\mathbf{p}}$. It is easy to check that the above Hamiltonian \eqref{eq:Hamil_dual} has a \textit{global} $\mathbb{Z}_N$ symmetry corresponding to $\left[\prod_{\mathbf{p}} e^{i \hat{\mathcal{B}}_{\mathbf{p}}}, \hat{H}_P  \right] = \left[\prod_{\mathbf{p}} e^{-i \hat{\mathcal{B}}_{\mathbf{p}}}, \hat{H}_P  \right] = 0$. In the limit $N \rightarrow \infty$, we get the global $U(1)$ symmetry.

We note here that the Hamiltonian \eqref{eq:Hamil_dual} involves every second plaquette of the original lattice. To simplify, we can coarse-grain the lattice by grouping two plaquettes, as illustrated in Fig.~\ref{fig:schem}(b). In this coarse-grained lattice, we consider the plaquette degrees of freedom to be located at the centers of these $2 \times 1$ squares (red circles in Fig.~\ref{fig:schem}(b)), describing a dual lattice to the original system.

To arrive at the $\mathbb{Z}_N$ gauge theory, we can invert the duality transformation by defining the electric-field operators on the link $(\mathbf{j}, \hat{x})$ between the sites $\mathbf{j}$ and  $\mathbf{j} + \hat{x}$ in the coarse-grained lattice as $\hat{\mathcal{E}}_{(\mathbf{j}, \hat{x})} = \hat{\mathcal{L}}_{\mathbf{p}} - \hat{\mathcal{L}}_{\mathbf{p} - \hat{y}}$ (see Fig.~\ref{fig:schem}(c)). Similarly, $\hat{\mathcal{E}}_{(\mathbf{j}, \hat{y})} = - \hat{\mathcal{L}}_{\mathbf{p}} + \hat{\mathcal{L}}_{\mathbf{p} - \hat{x}}$ on the link $(\mathbf{j}, \hat{y})$ between the site $\mathbf{j}$ and  $\mathbf{j} + \hat{y}$. On the other hand, $e^{i \hat{\mathcal{B}}_{\mathbf{p}}}$ can be redefined as $e^{i \hat{\mathcal{B}}_{\mathbf{p}}} \equiv \hat{U}_{(\mathbf{j}, \hat{x})} \hat{U}_{(\mathbf{j}+\hat{x}, \hat{y})} \hat{U}_{(\mathbf{j}+\hat{y}, \hat{x})}^{\dagger} \hat{U}_{(\mathbf{j}, \hat{y})}^{\dagger}$, where following commutation relations are maintained:
\begin{align}
    \left[\hat{\mathcal{E}}_{(\mathbf{j}, \hat{\delta})}, \hat{U}_{(\mathbf{j}, \hat{\delta})} \right] &= -\hat{U}_{(\mathbf{j}, \hat{\delta})}, 
    \nonumber \\
    \left[\hat{\mathcal{E}}_{(\mathbf{j}, \hat{\delta})}, \hat{U}_{(\mathbf{j}, \hat{\delta})}^{\dagger} \right] &= \hat{U}_{(\mathbf{j}, \hat{\delta})}^{\dagger}.
\end{align}
The Hamiltonian for the $\mathbb{Z}_N$ gauge theory, derived through the above duality transformation, is then written as
\begin{align}
    \hat{H}_G =& -\sum_{\mathbf{j}} \left( \hat{U}_{(\mathbf{j}, \hat{x})} \hat{U}_{(\mathbf{j}+\hat{x}, \hat{y})} \hat{U}_{(\mathbf{j}+\hat{y}, \hat{x})}^{\dagger} \hat{U}_{(\mathbf{j}, \hat{y})}^{\dagger} + \text{h.c.}\right) \nonumber \\
    & -2 g^2 \sum_{\mathbf{j}} \Bigg[ \cos\left(\frac{4\pi}{N} \hat{\mathcal{E}}_{(\mathbf{j}, \hat{x})}\right) \nonumber \\
    & + 2 \cos\left(\frac{2\pi}{N} \left(\hat{\mathcal{E}}_{(\mathbf{j}, \hat{x})} - \hat{\mathcal{E}}_{(\mathbf{j} + \hat{y}, \hat{x})}\right)\right) \nonumber \\
    & + \cos\left(\frac{2\pi}{N} \left(\hat{\mathcal{E}}_{(\mathbf{j}, \hat{y})} - \hat{\mathcal{E}}_{(\mathbf{j} + \hat{y}, \hat{y})}\right)\right)\Bigg].
\label{eq:Hamil_gauge}
\end{align}
Here the first term corresponds to the \textit{magnetic} plaquette term, while others are for interactions between the electric fields. The $\mathbb{Z}_N$ Gauss law corresponding to the local $\mathbb{Z}_N$ invariance is given by
\begin{equation}
    \sum_{\hat{\delta} = \hat{x}, \hat{y}}  \left(\hat{\mathcal{E}}_{(\mathbf{j}, \hat{\delta})} - \hat{\mathcal{E}}_{(\mathbf{j} - \hat{\delta}, \hat{\delta})} \right)= 0.
\end{equation}

In terms of standard $\mathbb{Z}_N$ operators $\hat{P}$ and $\hat{Q}$, defined as 
\begin{align}
    \hat{P} = \sum_{n=0}^{N-1} e^{i \frac{2 \pi}{N}n} \ket{n}\bra{n}, \quad
    \hat{Q} = \sum_{n=0}^{N-1} \ket{n}\bra{n + 1 \text{ mod } N},
\end{align}
and satisfying $\hat{Q} \hat{P} = e^{i \frac{2 \pi}{N}} \hat{P} \hat{Q}$, the Hamiltonian \eqref{eq:Hamil_dual} in the dual lattice and the  $\mathbb{Z}_N$ gauge theory Hamiltonian in Eq.~\eqref{eq:Hamil_gauge} can be rewritten as 
\begin{align}
    \hat{H}_P = -\sum_{\mathbf{p}} \Biggr[&\hat{P}_{\mathbf{p}} 
    + g^2 
    \biggr(\hat{Q}^2_{\mathbf{p}} \left(\hat{Q}^{\dagger}_{\mathbf{p}-\hat{y}}\right)^2 
 + 2 \, \hat{Q}^2_{\mathbf{p}} \hat{Q}^{\dagger}_{\mathbf{p}+\hat{y}} \hat{Q}^{\dagger}_{\mathbf{p}-\hat{y}} \nonumber \\
 &+ \hat{Q}_{\mathbf{p}} \hat{Q}^{\dagger}_{\mathbf{p}+\hat{x}} \hat{Q}^{\dagger}_{\mathbf{p}+\hat{y}}
\hat{Q}_{\mathbf{p}+\hat{x} + \hat{y}}\biggr)\Biggr]  + \text{h.c.},
\label{eq:Hamil_final}
\end{align}
\begin{align}
    \hat{H}_G = - & \sum_{\mathbf{j}} \Biggr[\hat{Q}_{(\mathbf{j}, \hat{x})} \hat{Q}_{(\mathbf{j}+\hat{x}, \hat{y})} \hat{Q}_{(\mathbf{j}+\hat{y}, \hat{x})}^{\dagger} \hat{Q}_{(\mathbf{j}, \hat{y})}^{\dagger} 
     \nonumber \\ 
     &+ g^2 \biggr( \hat{P}_{(\mathbf{j}, \hat{x})}^2
     + 2 \, \hat{P}_{(\mathbf{j}, \hat{x})}\hat{P}_{(\mathbf{j}+\hat{y}, \hat{x})}^{\dagger} + \hat{P}_{(\mathbf{j}, \hat{y})}\hat{P}_{(\mathbf{j}+\hat{y}, \hat{y})}^{\dagger} \biggr)\Biggr] \nonumber \\
     &+ \text{h.c.},
\label{eq:Hamil_gauge_final}
\end{align}
respectively. 
Here, in Eq.~\eqref{eq:Hamil_final}, we identify $\hat{P}_{\mathbf{p}} \equiv e^{i \hat{\mathcal{B}}_{\mathbf{p}}}$ and $\hat{Q}_{\mathbf{p}} \equiv e^{i \frac{2\pi}{N}\hat{\mathcal{L}}_{\mathbf{p}}}$, while in Eq.~\eqref{eq:Hamil_gauge_final} we use $\hat{P}_{(\mathbf{j}, \hat{\delta})} \equiv e^{i \frac{2\pi}{N}\hat{\mathcal{E}}_{(\mathbf{j}, \hat{\delta})}}$ and 
$\hat{Q}_{(\mathbf{j}, \hat{\delta})} \equiv  \hat{U}_{(\mathbf{j}, \hat{\delta})}$.
Unless stated otherwise, in the rest of the paper we will consider Hamiltonian $\hat{H}_P$ (i.e., Eq.~\eqref{eq:Hamil_final}) in the dual lattice that possesses global $\mathbb{Z}_N$ symmetry to characterize the properties of the system in the low-energy manifold. We should note that the duality relation between the Hamiltonians \eqref{eq:Hamil_final} and \eqref{eq:Hamil_gauge_final} is the generalization of well-known Wegner duality defined between 2D quantum Ising model and $\mathbb{Z}_2$ gauge theory~\cite{Wegner1971, Wegner2014} (see also~\cite{Horn1979}).

In case of standard $\mathbb{Z}_N$ gauge theories with finite $N$, the phase diagram consists of two gapped phases -- a topologically ordered deconfined phase at small values of coupling strength $g$ and a confined phase at large values of $g$~\cite{Kogut1975, Drell1979, Banks1977}, where following Polyakov's prediction the former disappears at the $U(1)$ limit of $N \rightarrow \infty$~\cite{Polyakov1975,Polyakov1977}. In the present scenario, the confined phase is a symmetry-broken phase due to the unique nature of the interactions (see~\cite{Dutta2017}). Moreover, using a mean-field argument~\cite{Dutta2017}, it has been conjectured that a gapless Bose liquid phase appears at an `intermediate' values of coupling strength $g$, where the population imbalance of the $b$-bosons on plaquettes (given by non-zero $\hat{\mathcal{B}}_{\mathbf{p}}$ i.e., the magnetic charge) propagates freely along the $y$-direction. Consequently, the system in this intermediate phase can be effectively described as a series of one-dimensional gapless phases stacked together in the $x$-direction.
On the other hand, another gapless phase of dipolar nature has been conjectured to appear for $N \rightarrow \infty$ and $g \rightarrow 0$ with $g N$ being finite.

In this work, we provide robust numerical validation of these predictions for finite $N$ by employing density matrix renormalization group (DMRG) algorithm~\cite{White1992, White1993} with matrix-product state (MPS)\cite{Schollwoeck2011, Orus2014, Ran2020} ansatz. This approach is applied to quasi-2D ladder and cylinder systems that are infinitely extended in the $y$-direction but have a finite length or circumference $L_x$ in the $x$-direction. Specifically, we use the infinite-size version of DMRG (iDMRG)\cite{McCulloch2008}, where the underlying infinite MPS (iMPS) extends infinitely in the $y$-direction and wraps around the cylinder in a zigzag pattern along the $x$-direction. Moreover, we make use of $\mathbb{Z}_N$ symmetry preserving tensors that conserve the global $\mathbb{Z}_N$ symmetry (corresponding to $\left[\prod_{\mathbf{p}} \hat{P}^{(\dagger)}_{\mathbf{p}}, \hat{H}_P  \right] = 0$) throughout the calculations.

\section{The Phase diagram}
\label{sec:phase}

In this section, we systematically analyze the phase diagram of Hamiltonian~\eqref{eq:Hamil_final}
 by examining three different geometries: (1) the 1D case with $L_x = 1$, (2) a two-leg ladder geometry with $L_x = 2$, and (3) cylinder geometries with $L_x > 2$ and periodic boundary conditions in the $x$-direction. For this purpose, we consider two quantities, namely the von Neumann entanglement entropy of the half-infinite block and the iMPS correlation length for a given iMPS bond dimension $\chi$. Such technique is called finite-entanglement scaling and was introduced in a series of publications \cite{Nishino1996, Tagliacozzo2008, Pollmann2009, Pirvu2012, Kjall2013}.

 The von Neumann entanglement entropy of the half-infinite block for a given bond dimension $\chi$ is defined as
 \begin{equation}
      S_{\chi} = - \text{Tr} \left[ \rho^L_{\chi} \ln(\rho^L_{\chi}) \right] = - \sum_k \lambda_k^2 \ln \lambda_k^2,
 \end{equation}
where $\rho^L_{\chi} = \text{Tr}_{R} \left[\ket{\psi}\bra{\psi}\right]$ represents the reduced density matrix of the left half of the system obtained by tracing out the right infinite half of the system, and $\{\lambda_k\}$ are the corresponding Schmidt coefficients. For an iMPS with bond dimension $\chi$, the correlation length is given by $\xi_{\chi} = -1/\ln|\epsilon_2|$, where $\epsilon_2$ is the second largest eigenvalue of the iMPS transfer matrix~\cite{Kjall2013}. In gapless systems, where the correlation length diverges, $\xi_{\chi}$ represents an artificially introduced length scale due to the finite iMPS bond dimension $\chi$. Typically, $\xi_{\chi} \sim \chi^{\beta}$, with $\beta$ being a scaling exponent.

\subsection{One-dimensional scenario}

\begin{figure}
\includegraphics[width=\linewidth]{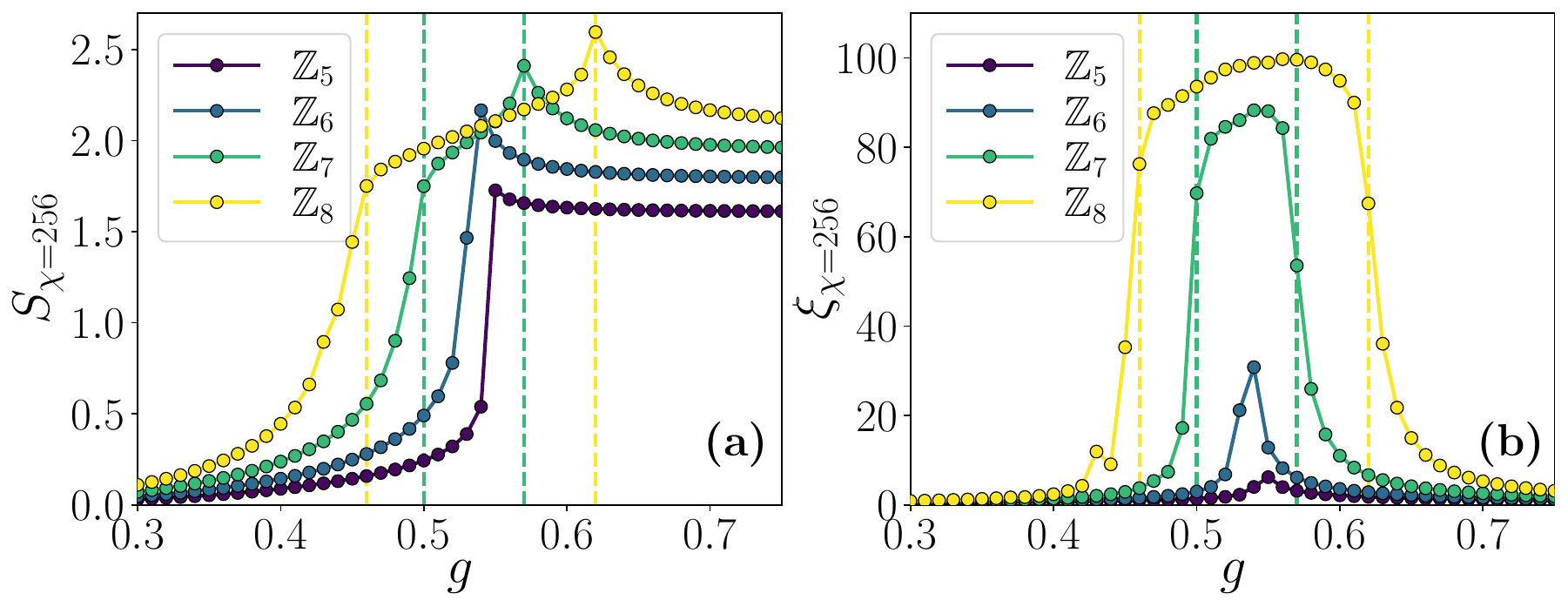}
\caption{(Color online.)
    (a) The entanglement entropy $S_{\chi}$ and (b) the correlation length $\xi_{\chi}$ for the ground state of the $\mathbb{Z}_N$-symmetric 1D system (Eq.~\eqref{eq:Hamil_1D}) as a function of $g$, calculated for $N=5,6,7,8$ with iMPS bond dimension $\chi=256$. Only two gapped phases are observed for $N \le 6$, while a gapless phase emerges at $N = 7$ around $g=0.5$ and becomes more robust at $N=8$. Vertical dashed lines indicate the gapless phases for $N=7$ and $8$.
}
\label{fig:1d}
\end{figure}

In 1D geometry, the last term in the Hamiltonian~\eqref{eq:Hamil_final}, i.e., the four-plaquette term, is absent, and the system is then described by the Hamiltonian:
\begin{align}
    \hat{H}_P^{1D} =& -\sum_{\mathbf{p}} \left[\hat{P}_{\mathbf{p}} 
    + g^2 
    \left(\hat{Q}^2_{\mathbf{p}} \left(\hat{Q}^{\dagger}_{\mathbf{p}-\hat{y}}\right)^2 
 + 2 \, \hat{Q}^2_{\mathbf{p}} \hat{Q}^{\dagger}_{\mathbf{p}+\hat{y}} \hat{Q}^{\dagger}_{\mathbf{p}-\hat{y}} \right)\right]  \nonumber \\ &+ \text{h.c.}
 \label{eq:Hamil_1D}
\end{align}
In terms of gauge theory, such a 1D system of plaquettes in the dual lattice corresponds to a local $\mathbb{Z}_N$ invariant system in a two-leg ladder geometry described by Hamiltonian~\eqref{eq:Hamil_gauge_final} where the last term, i.e., the $\hat{P}_{(\mathbf{j}, \hat{y})}\hat{P}_{(\mathbf{j}+\hat{y}, \hat{y})}^{\dagger}$ term, is absent.

First, we observe that in absence of the three-plaquette term (i.e., the last term) in Eq.\eqref{eq:Hamil_1D}, the odd and even $\mathbb{Z}_N$ sectors decouple for even $N$ due to the squared $\hat{Q}^2_{\mathbf{p}}$ operators present in the second term. This decoupling maps the system into two copies of a $\mathbb{Z}_{N/2}$ symmetric $N/2$-state quantum Clock model~\cite{Einhorn1980, Radicevic2018}. It has become clear that for the $q$-state quantum Clock model, while there are two gapped phases for $q \leq 4$, an intermediate gapless phase appears for $q \geq 5~$\cite{Ortiz2012, Bingnan2020}. In our scenario, we first confirm (results not shown) that the system described by Eq.~\eqref{eq:Hamil_1D}, without the last term, exhibits an intermediate gapless phase for $N \geq 10$, regardless of whether $N$ is even or odd.

The presence of the three-plaquette term in Eq.~\eqref{eq:Hamil_1D} introduces coupling between the odd and even sectors, preventing the system from being mapped into two independent copies of the $N/2$-state quantum clock model. This coupling fundamentally alters the structure and behavior of the system, raising a critical question: does the gapless intermediate phase, observed in the absence of the three-plaquette term, persist when this term is included? If so, at what value of $N$ does this phase begin to appear? To address this, we carefully examine the phase diagram of the system in the presence of the three-plaquette interaction as elaborated below.

Figure~\ref{fig:1d} shows the entanglement entropy and the correlation length for the ground state of Hamiltonian~\eqref{eq:Hamil_1D} for different values of $N$ calculated for iMPS bond dimension $\chi=256$. Both the entanglement entropy and the correlation length approach to zero in the limit $g \rightarrow 0$ for any values of $N \leq 8$. Therefore, we detect a (trivial) gapped phase in the small $g$ limit. In the dual picture of $\mathbb{Z}_N$ gauge theory, this gapped phase corresponds to the deconfined phase with non-trivial topological order~\cite{Kitaev2003, Giudice2022, Simon2023}. The non-trivial topological character of this deconfined phase is verified by checking the degeneracies in the entanglement spectrum~\cite{Pollmann2010} of the ground state of Hamiltonian~\eqref{eq:Hamil_gauge_final} in the small $g$ limit.

The correlation length approaches to zero also in the high $g$ limit indicating the presence of another gapped phase. This gapped phase at strong coupling corresponds to a spontaneous symmetry broken (SSB) phase where global $\mathbb{Z}_N$ symmetry of Hamiltonian~\eqref{eq:Hamil_1D} gets broken. That is why the entanglement entropy saturates to $S \rightarrow \ln N$ for $g \rightarrow \infty$ as in our calculation iMPS ansatz preserves the global $\mathbb{Z}_N$ symmetry. In case of symmetry-broken states, the order parameter $\sum_{\mathbf{p}} \braket{\hat{Q}_{\mathbf{p}} + \hat{Q}^{\dagger}_{\mathbf{p}}}$ becomes non-zero in this large-$g$ phase. Within our $\mathbb{Z}_N$-symmetry preserving iMPS calculations, this symmetry-breaking can be verified by checking $|\braket{\hat{Q}_{\mathbf{p}} \hat{Q}^{\dagger}_{\mathbf{p} + R\hat{y}}}| \approx 1$ for large $R$. In the dual gauge-invariant system, this strong coupling phase corresponds to an SSB confined phase with $\sum_{\mathbf{j}, \hat{\delta}} \braket{\hat{P}_{(\mathbf{j}, \hat{\delta})} + \hat{P}_{(\mathbf{j}, \hat{\delta})}^{\dagger}}$ being the order parameter. Moreover, we notice that there is only one cusp (maxima) in the profile of entropy (correlation length) for $N \leq 6$, indicating a direct phase transition between these gapped phases.

Starting from $N=7$, the phase diagram undergoes a notable change (Fig.~\ref{fig:1d}) with the emergence of an additional phase at intermediate values of $g$\footnote{It should be noted that we observe a robust gapless phase starting from $N=7$. Determining whether the gapless phase also exists for $N=6$ or even $N=5$ would require a very fine-grained grid along the $g$-axis, significantly increasing computational complexity. Therefore, we refrain from conducting such an analysis.}. The high correlation lengths observed in this intermediate phase suggest its critical nature. Therefore, the inclusion of the three-plaquette term causes the gapless phase to appear at a lower value of $N$ compared to its absence.  
Since gapless phases in 1D are essentially Luttinger liquids~\cite{Giamarchi2003, Gogolin2004}, we anticipate the same in this system. In the following Sec.~\ref{sec:characterize}, we validate this expectation by examining the scaling of entanglement entropy. In the dual $\mathbb{Z}_N$ gauge theory in two-leg ladder geometry, this gapless phase corresponds to a gapless deconfined phase.
For larger values of $N$, we find that the region of this gapless phase increases, which indicates the robustness of this phase with respect to increasing $N$. We should note that the origin of this gapless phase is fundamentally different from the Coulomb phase recently observed in $\mathbb{Z}_N$ lattice gauge theories in the presence of Higgs matter in two-leg ladder geometry~\cite{Nyhegn2021}, as such Coulomb phase disappears for pure gauge theory, i.e., in absence of the Higgs matter.

\begin{figure*}
   \includegraphics[width=0.667\linewidth]{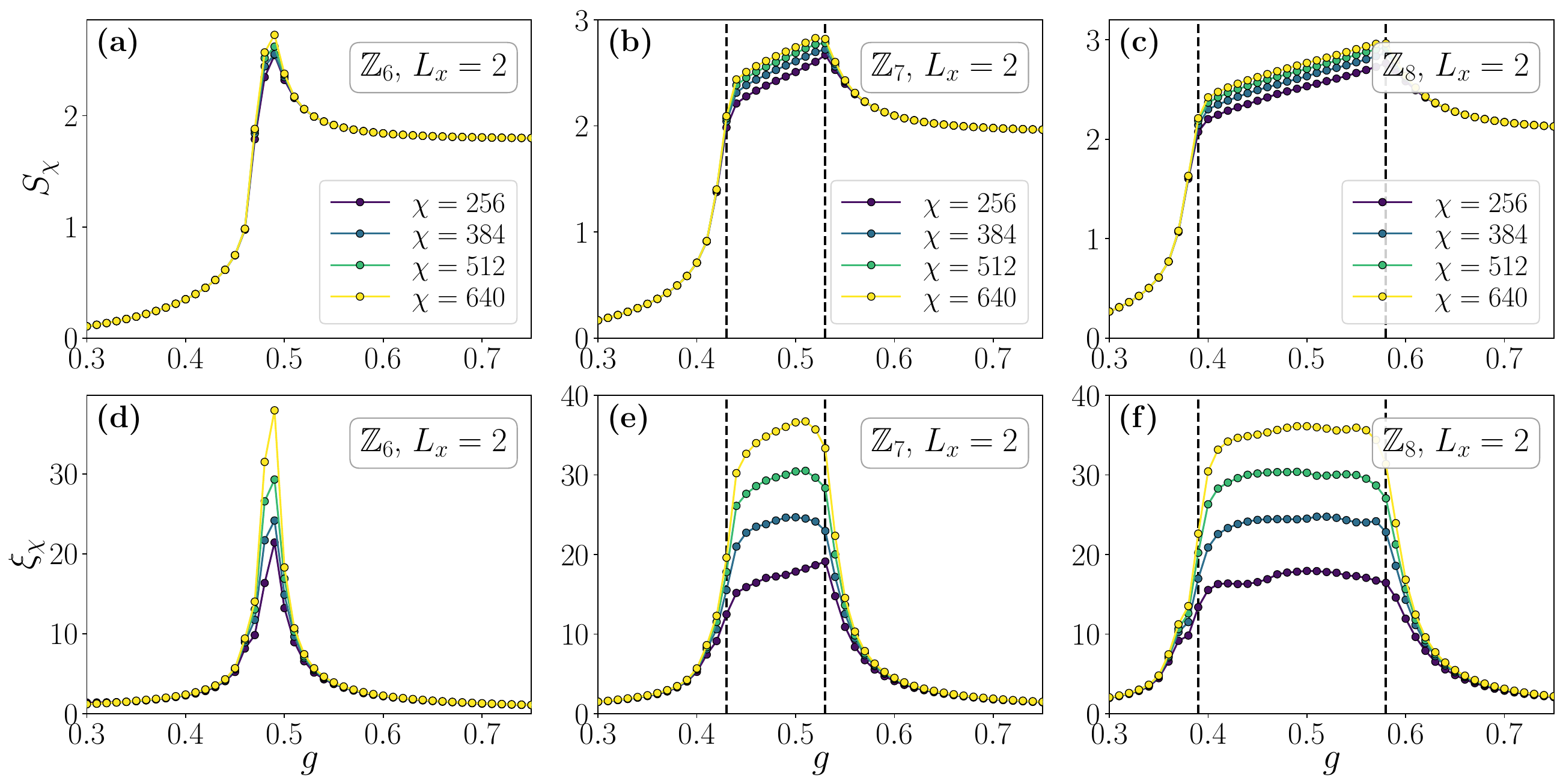}
    \caption{ (Color online.)
    The entanglement entropy $S_{\chi}$ and the correlation length $\xi_{\chi}$ of the ground state of the system~\eqref{eq:Hamil_final} in a two-leg ladder geometry as a function of $g$, calculated for $N=6, 7$ and $8$. The data indicates the presence of two gapped phases for $N=6$. At $N=7$ the gapless phase emerges in the area $0.43 \lesssim g \lesssim 0.53$, while at $N=8$ it encompasses a wider region $0.39 \lesssim g \lesssim 0.58$. The black vertical lines demarcate the gapless region.
    }\label{fig:ladder}
\end{figure*}

\subsection{Two-leg ladder system}

In this subsection, we consider the simplest extension beyond 1D: a two-leg ladder geometry (i.e., $L_x = 2$), which includes the four-plaquette term (the last term in Eq.\eqref{eq:Hamil_final}). In this setup, the system maps to a three-leg ladder system with open boundary conditions along the $x$-direction in the dual $\mathbb{Z}_N$ gauge theory (Eq.\eqref{eq:Hamil_gauge_final}). The key question we aim to address here is whether the conclusions drawn in the previous subsection still hold as we make the transition from a 1D to a quasi-2D system. While we expect the gapped phases to persist, the fate of the gapless phase remains uncertain. The four-plaquette term may act as a relevant perturbation, potentially causing a gap to open.

\begin{figure}
    \includegraphics[width=\linewidth]{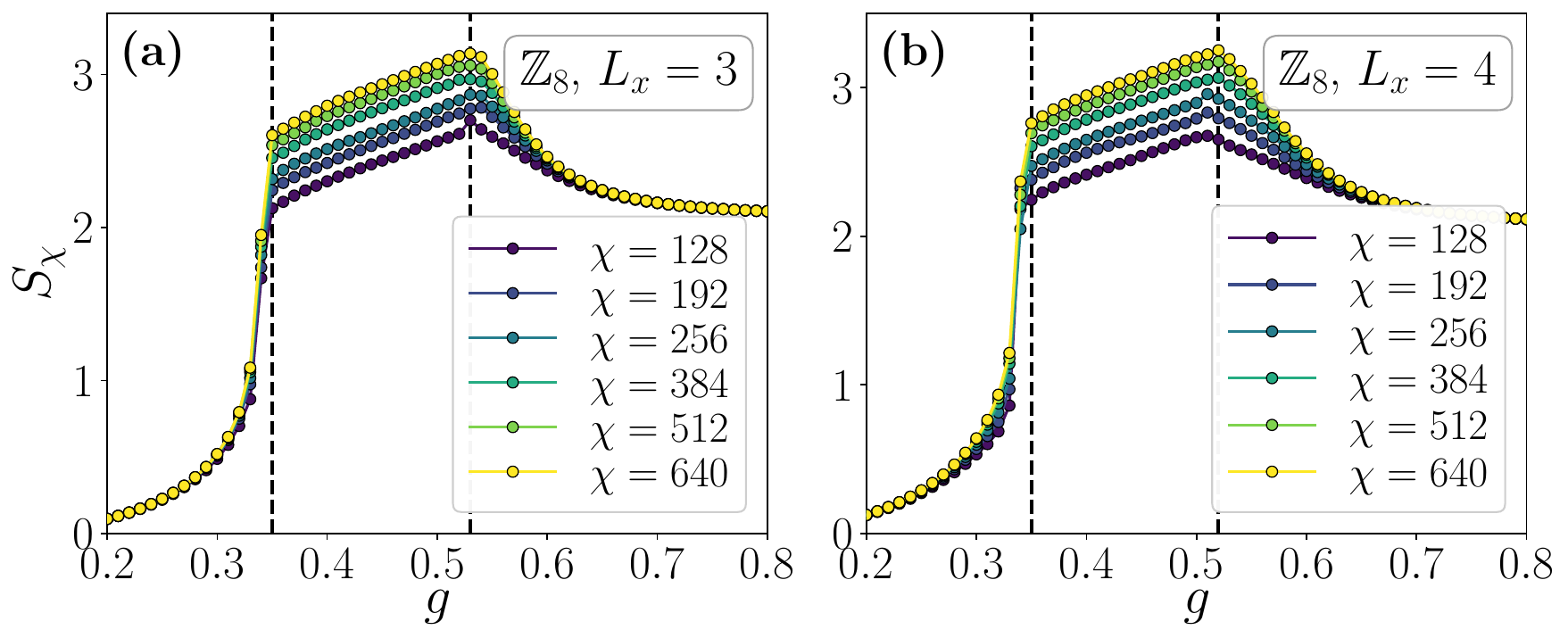}
    \caption{ (Color online.)
    The entanglement entropy $S_{\chi}$ of the ground state of the $\mathbb{Z}_8$-symmetric system~\eqref{eq:Hamil_final} embedded in cylinders with transverse dimensions $L_x=3$ and $L_x=4$ as a function of $g$. Similar to the situations for $L_x=1,2$, we observe two gapped phases, confined and deconfined for large and small values of $g$ respectively, and a gapless phase sandwiched between these two gapped phases at intermediate values of $g$. 
    The gapless phase persists in the range $0.35 \lesssim g \lesssim 0.53$ for $L_x=3$ and $0.35 \lesssim g \lesssim 0.52$ for $L_x=4$.
    }
    \label{fig:cylinder}
\end{figure}

In Fig.~\ref{fig:ladder}, we show the entanglement entropy $S_{\chi}$ and the correlation length $\xi_{\chi}$ for the ladder system for different values of $N$ and iMPS bond dimension $\chi$ as depicted in the figure. Both the entropy and the correlation length show similar features as in the 1D scenario. While for $N \leq 6$ the system exhibits only two gapped phases (the deconfined and confined phases in the language of $\mathbb{Z}_N$ gauge theory)\footnote{We stress again that there is a possibility of having a minuscule gapless region sandwiched between the confined and deconfined regions for $N=6, 5$. However, distinguishing this phase from a direct critical phase transition between the confined and deconfined phases is very challenging with iDMRG simulations with finite bond dimension and finite grid-spacing in~$g$.}, we observe extended gapless region for $N \geq 7$ in the intermediate values of $g$. By monitoring large values of correlation length that increases with increasing iMPS bond dimension, we identify the gapless phase to be within the range $0.43 \lesssim g \lesssim 0.53$ for $N=7$ and $0.39 \lesssim g \lesssim 0.58$ for $N=8$. We should note that nature of this gapless phase can be described by coupled pillars of Luttinger liquids, as the four-plaquette term now couples the two chains (along $y$ direction) of former Luttinger liquids along the $x$-direction. We leave this discussion for Sec.~\ref{sec:characterize}.

\subsection{Three- and four-legged cylinder system}

Next, we analyze the fate of the intermediate gapless phase by increasing the transverse dimension $L_x$. Specifically, we consider the Hamiltonian~\eqref{eq:Hamil_final} in three- and four-legged cylinder geometries, i.e., with periodic boundary condition along $x$ axis with $L_x=3$ and $L_x=4$. In the dual lattice, such systems translate to $\mathbb{Z}_N$ gauge theory (i.e., Eq.~\eqref{eq:Hamil_gauge_final}) embedded in cylinders with transverse dimension $L_x=3$ and $L_x=4$ respectively.

The numerical result for the entanglement entropy $S_{\chi}$ for the ground state of $\mathbb{Z}_8$-symmetric Hamiltonian~\eqref{eq:Hamil_final} for different iMPS bond dimension $\chi$ is given in Fig.~\ref{fig:cylinder}. The qualitative scenario remains effectively unchanged from the previous cases. From the cusps in the entropy profile, we detect the gapless region in the range $0.35 \lesssim g \lesssim 0.53$ for $N=7$ and $0.35 \lesssim g \lesssim 0.52$ for $N=8$. It is therefore evident that the gapless region remains robust at larger transverse dimension $L_x$, and we expect it to remain true even for fully extended 2D case.

\begin{table*}
\begin{tabular}{|c|c|c|c|}
\hline
System & At small $g$ & At intermediate $g$ ($N \geq 7$) & At large $g$ \\
\hline
$\hat{H}_{P}$ (Eq.~\eqref{eq:Hamil_final})  & Trivial gapped phase & Gapless 1D liquids & SSB phase \\
(Global $\mathbb{Z}_N$ symmetry) & & (Weekly coupled Luttinger liquids) & $\sum_{\mathbf{p}} \braket{\hat{Q}_{\mathbf{p}} + \hat{Q}^{\dagger}_{\mathbf{p}}} \neq 0$ \\
\hline 
$\hat{H}_{G}$ (Eq.~\eqref{eq:Hamil_gauge_final}) & Deconfined phase & Gapless deconfined phase & SSB confined phase \\
(Local $\mathbb{Z}_N$ symmetry) & (Topologically ordered) & & $\sum_{\mathbf{j}, \hat{\delta}} \braket{\hat{P}_{(\mathbf{j}, \hat{\delta})} + \hat{P}_{(\mathbf{j}, \hat{\delta})}^{\dagger}} \neq 0$ \\
\hline
\end{tabular}
\caption{Different phases found in our study for cylinder/ladder systems with width $L_x \leq 4$.}
\label{tab:phases}
\end{table*}

The summary of the phase diagram found in our calculation is presented in the Table~\ref{tab:phases}. It should be noted that contrary to the predictions of~\cite{Dutta2017}, we do not find a gapless at $g \rightarrow 0$ predicted for large $N$. This is, probably, because we have limited our analysis for finite $N$~\footnote{We note that we have verified this by calculating the ground state of the system till $N=15$ for a few small values of $g$.}. Increasing $N$ to very large numbers would increase the computational cost due to the increment in the local Hilbert space dimension. The absence of this gapless phase at low coupling may also be attributed to the limited cylinder width which we consider with MPS ansatz.

\section{Characterization of the gapless deconfined phases}
\label{sec:characterize}

In this section, we first characterize the gapless deconfined phases appearing for $N \geq 7$ by scrutinizing different correlation functions. Next, we determine the central
charge of the underlying conformal field theory (CFT) in these critical regions from the scaling of entanglement entropy.

\begin{figure}[htb]
\includegraphics[width=\linewidth]{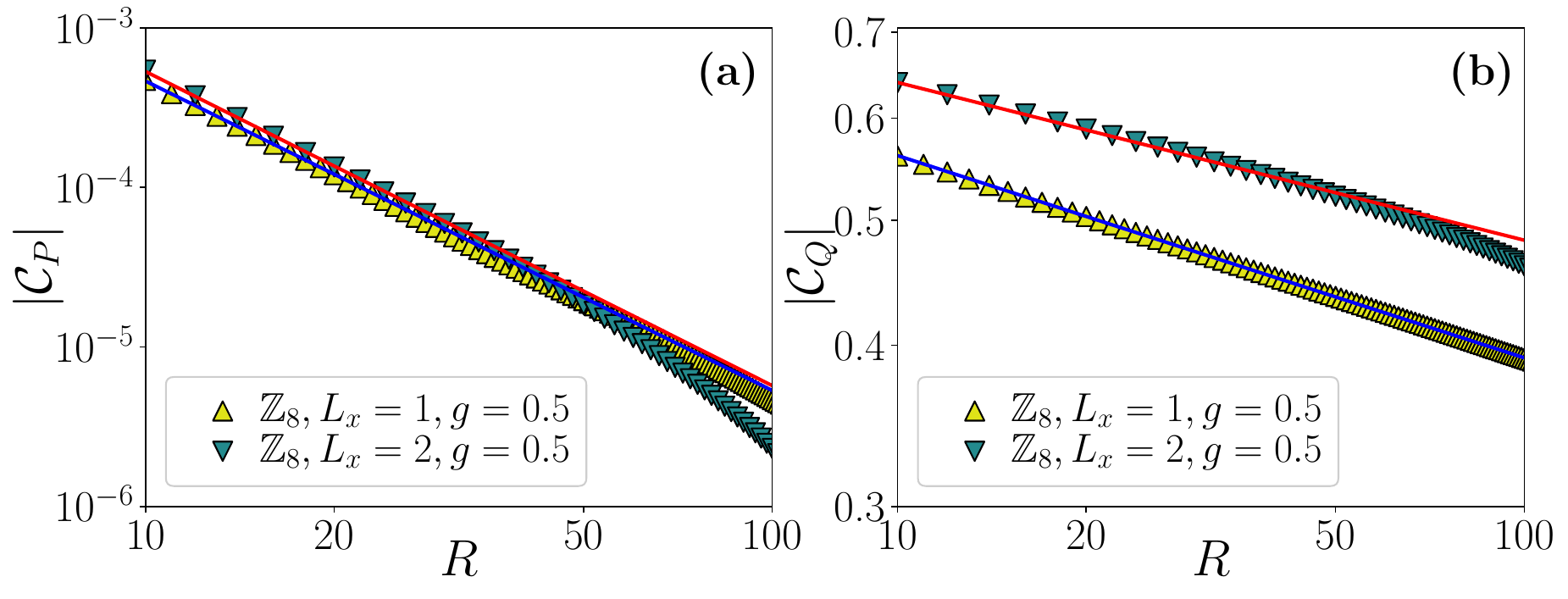}
\includegraphics[width=\linewidth]{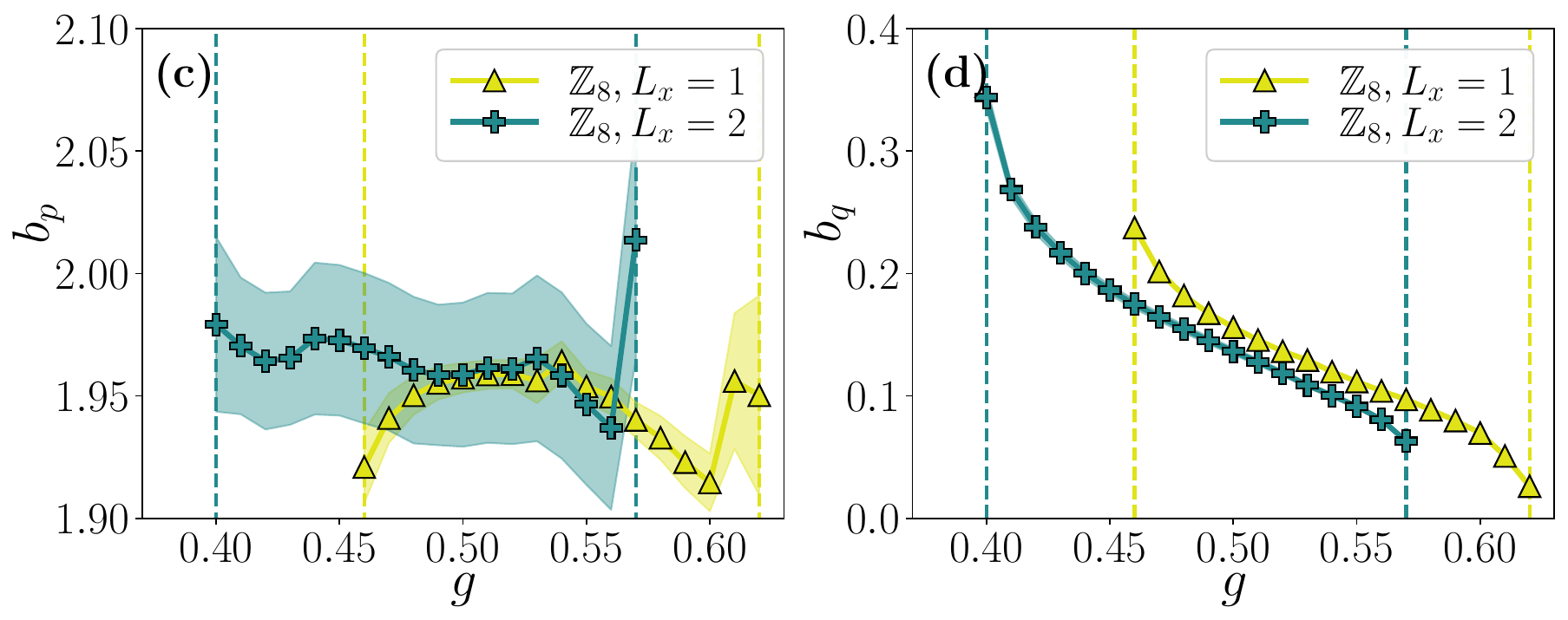}
\caption{(Color online.) (a)-(b) The power-law decay of the correlation functions $\mathcal{C}_P$ and $\mathcal{C}_Q$ inside the gapless phase for $N=8$ and $L_x=1, 2$. We set iMPS bond dimension $\chi$ to 512 and 1280 respectively for $L_x=1$ and $L_x=2$. The deviation from power-law decay seen for $L_x=2$ at large $R$ is due the finite correlation length $\xi_{\chi}$ imposed by finite $\chi$. Both $x$ and $y$ axes are in log-scale.
(c)-(d) The power-law exponents $b_p$ and $b_q$, extracted by fitting the $\mathcal{C}_P$ and $\mathcal{C}_Q$ data respectively, in the entire gapless region for $N=8$ and $L_x=1, 2$. The shaded regions indicate the error bars and the vertical dashed lines depict the extent of the gapless regions. Here, we set iMPS bond dimension $\chi$ to 512 and 640 respectively for $L_x=1$ and $L_x=2$.}
\label{fig:corr}
\end{figure}

\subsection{Correlation functions}

We consider the following two-point connected correlation functions along the $y$-direction:
\begin{align}
    \mathcal{C}_P(R) := \braket{\hat{P}^{\dagger}_{\mathbf{p}} \hat{P}_{\mathbf{p} + R \hat{y}}} - \braket{\hat{P}^{\dagger}_{\mathbf{p}}} \braket{\hat{P}_{\mathbf{p} + R \hat{y}}}, \nonumber \\
    \mathcal{C}_Q(R) := \braket{\hat{Q}^{\dagger}_{\mathbf{p}} \hat{Q}_{\mathbf{p} + R \hat{y}}} - \braket{\hat{Q}^{\dagger}_{\mathbf{p}}} \braket{\hat{Q}_{\mathbf{p} + R \hat{y}}}.
\end{align}
In our calculations, since we are employing $\mathbb{Z}_N$ symmetry preserving iMPS ansatz, the local expectation values $\braket{\hat{Q}^{\dagger}_{\mathbf{p}}}$ or $\braket{\hat{Q}_{\mathbf{p}}}$ necessarily vanishes.

At the small-$g$ gapped phase, both the correlations decay exponentially with $R$ as it is expected for a trivial gapped phase. On the other hand, at the strong-coupling gapped phase, $\mathcal{C}_P$ decays to zero exponentially, while $|\mathcal{C}_Q|$ saturates to one for large enough $R$ due to SSB. 
For the gapless phase at intermediate $g$, we expect both the correlations to decay algebraically to zero: $|\mathcal{C}_P(R)| \sim R^{-b_p}$ and $|\mathcal{C}_Q(R)| \sim R^{-b_q}$ as long as $R \lesssim \xi_{\chi}$. 
Figure~\ref{fig:corr}(a) and (b) display the power-law behavior of these correlators inside the gapless region for $N=8$ and $L_x = 1$ and $2$. By extracting the exponent $b_p$ for the correlator $\mathcal{C}_P$ (Fig.~\ref{fig:corr}(c)) we find that $b_p \approx 2$, i.e., $|\mathcal{C}_P(R)| \sim 1/R^2$, irrespective of the values of $g$. In the case of $\mathcal{C}_Q$ the decay is much slower where $b_q$ varies with varying $g$ (Fig.~\ref{fig:corr}). This slow decay of $\mathcal{C}_{Q}$ is expected as this gapless phase is adjacent to the SSB phase where $b_q \rightarrow 0$.

\subsection{Scaling of entanglement entropy}

To understand the nature of the gapless phase, we consider the scaling of entanglement entropy with respect to the correlation length. For 1D (or quasi-2D, as in the present scenario) critical systems, the entanglement entropy increases logarithmically with the correlation length following the well-known relation~\cite{callan_geometric_1994, vidal_PRL_2003, calabrese_entanglement_2004}:
\begin{equation}
    S_{\chi}=\frac{c}{6} \log \xi_{\chi} + b',
    \label{eq:entlog}
\end{equation}
where $c$ is the central charge of the underlying CFT and $b'$ is a non-universal constant.

\begin{figure}[htb]
\includegraphics[width=\linewidth]{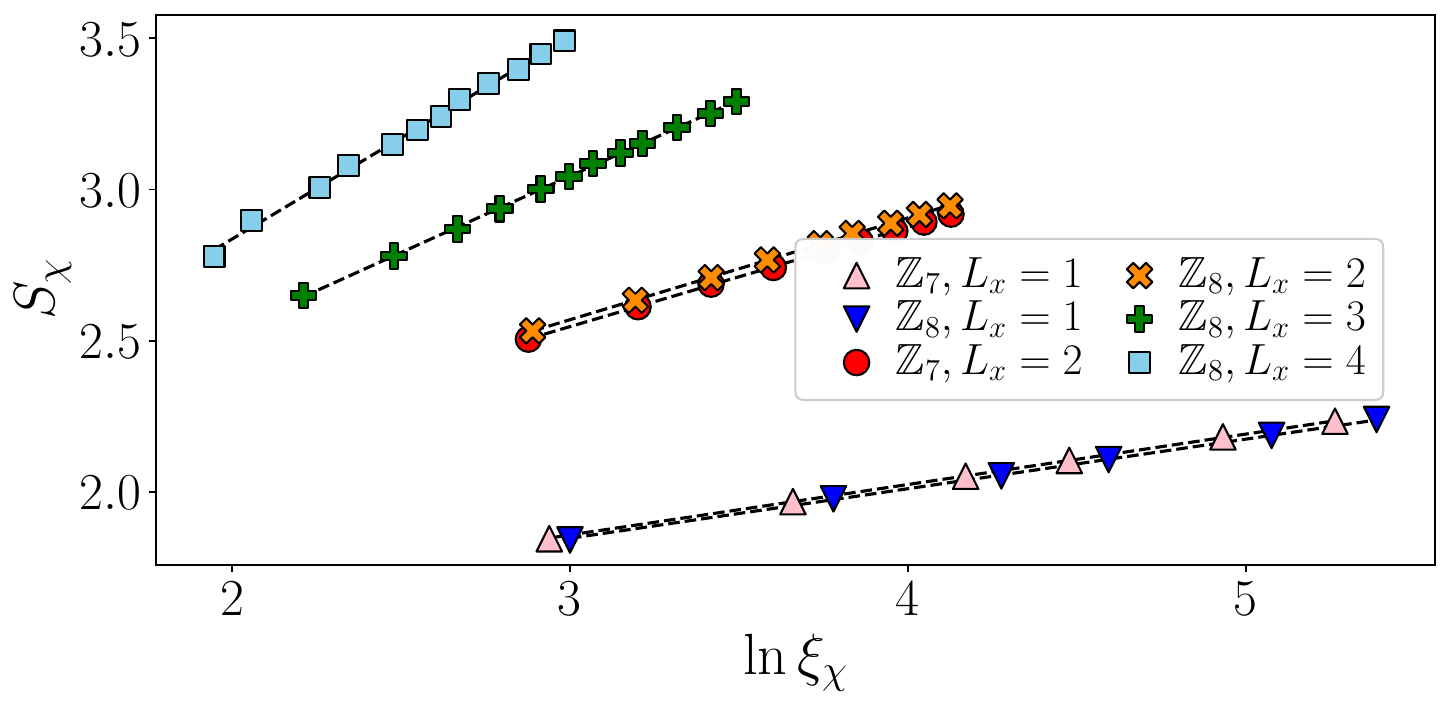}
\caption{(Color online.) The scaling of entanglement entropy $S_{\chi}$ as a function of the correlation length $\xi_{\chi}$ according to Eq.~\eqref{eq:entlog} for different system-width $L_x$ and $N$. The extracted values of the central charge, along with other details, are given in Table~\ref{tab:central_charge}.
}
\label{fig:central_charge}
\end{figure}

By fitting Eq.\eqref{eq:entlog} to the entropic data for various iMPS bond dimensions within the gapless phase (see Fig.\ref{fig:central_charge}), we find the central charge $c = 1$ for $L_x = 1$ in both $\mathbb{Z}_7$ and $\mathbb{Z}_8$ symmetric systems. This result aligns with expectations for 1D gapless phases described by Luttinger liquid theory. For $L_x \geq 2$, the four-plaquette term in Eq.\eqref{eq:Hamil_final} couples $L_x$ identical copies of such Luttinger liquids. In principle, this coupling could be relevant and potentially open a (partial) gap in the system, resulting in $c \leq L_x$. However, the scaling analysis presented in Fig.\ref{fig:central_charge} shows $c = L_x$ up to small numerical errors. This indicates that the four-plaquette interaction is an irrelevant perturbation, and following $c$-theorem~\cite{Zamolodchikov1986}, the system essentially consists of $L_x$ decoupled Luttinger liquids in the long-wavelength (low-energy, infrared) limit. Therefore, the gapless phase at intermediate $g$ effectively remains a set of pure 1D critical states, which we expect to hold even for large $L_x$ and in a fully extended 2D setting.

The extracted numerical values of $c$ along with other system parameters, considered for the plot of Fig.~\ref{fig:central_charge}, are presented in the Table~\ref{tab:central_charge}.

\begin{table}[htb]
    \centering
    \begin{tabular}{|c|c|c|c|c|}
    \hline
    \ $N$ \ & \ $L_x$ \ & \ $g$ \ & $\chi$-range considered & $c$ \\
    \hline
    7 & 1 & 0.55 & [64, 512] & 1.00(2) \\ 
    8 & 1 & 0.55 & [64, 512] & 0.98(2) \\
    7 & 2 & 0.5 & [256, 1280] & 1.98(2) \\
    8 & 2 & 0.5 & [256, 1280] & 2.02(3) \\
    8 & 3 & 0.45 & [256, 2048] & 3.03(4) \\
    8 & 4 & 0.45 & [256, 2048] & 4.0(1) \\
    \hline
    \end{tabular}
    \caption{The numerically extracted values of the central charge $c$ by fitting the entropic data to Eq.~\eqref{eq:entlog} for different system (and iMPS) parameters. The values of $g$ are chosen in such a way that the system remains within the gapless region.}
    \label{tab:central_charge}
\end{table}

\section{Conclusion}
\label{sec:conclu}

Ultracold atoms in particular optical lattice potentials allow for the creation of models with local symmetries. In the current work, we have considered a specific optical lattice setup involving two species of ultracold bosons. In the low-energy manifold, inter-species dynamics give rise to unique Abelian $\mathbb{Z}_N$ gauge theories. By using infinite-system density matrix renormalization group simulations, 
we have shown that such quantum systems exhibit particularly interesting phases of matter. While for $N \leq 6$ only two gapped phases exist as in standard $\mathbb{Z}_N$ gauge theories, we confirm the presence of a gapless (deconfined) phase for $N \ge 7$ as well as its robustness with respect to increasing $N$. This gapless phase exhibits the same conformal critical properties as the one-dimensional Bose liquid phase, in line with field theoretical predictions~\cite{Dutta2017}, where its description is same as the collection of weakly coupled chains of identical Luttinger liquids. 

On the other hand,  we cannot confirm the existence of another gapless (presumably dipolar liquid) phase. This is probably a consequence of 
finite $N$ considered here or the physical geometry of the system we can explore, in which one of the two dimensions is bound to be very short. As a result, even if we cannot observe it, we cannot exclude the possibility of its existence in the thermodynamic limit of the  U(1) symmetric theory where the dipolar liquid phase is conjectured to exist, i.e.,  in the limit  ($L_x\to \infty, g \to 0$, $N \to \infty$, $gN = {\rm const}$). Numerical calculations at large values of $N$ are beyond the scope of the current manuscript.

\acknowledgments
The work of M.V.R. was funded by the National Science Centre, Poland, under the OPUS call within the WEAVE program 2021/43/I/ST3/01142.
L.T. acknowledges support from the Proyecto Sinérgico CAM Y2020/TCS-6545 NanoQuCo-CM, the CSIC Research Platform on Quantum Technologies PTI-001, and from the Grant TED2021-130552B-C22 funded by MCIN/AEI/10.13039/501100011033 and by the ``European Union NextGenerationEU/PRTR'',
and Grant PID2021-127968NB-I00 funded by MCIN/AEI/10.13039/501100011033.
M.L. acknowledges support from: European Research Council AdG NOQIA; MCIN/AEI (PGC2018-0910.13039/501100011033, CEX2019-000910-S/10.13039/501100011033, Plan National FIDEUA PID2019-106901GB-I00, Plan National STAMEENA PID2022-139099NB, I00, project funded by MCIN/AEI/10.13039/501100011033 and by the “European Union NextGenerationEU/PRTR" (PRTR-C17.I1), FPI); QUANTERA DYNAMITE PCI2022-132919, QuantERA II Programme co-funded by European Union’s Horizon 2020 program under Grant Agreement No 101017733); Ministry for Digital Transformation and of Civil Service of the Spanish Government through the QUANTUM ENIA project call - Quantum Spain project, and by the European Union through the Recovery, Transformation and Resilience Plan - NextGenerationEU within the framework of the Digital Spain 2026 Agenda; Fundació Cellex; Fundació Mir-Puig; Generalitat de Catalunya (European Social Fund FEDER and CERCA program, AGAUR Grant No. 2021 SGR 01452, QuantumCAT \ U16-011424, co-funded by ERDF Operational Program of Catalonia 2014-2020); Barcelona Supercomputing Center MareNostrum (FI-2023-3-0024), HORIZON-CL4-2022-QUANTUM-02-SGA PASQuanS2.1, 101113690, EU Horizon 2020 FET-OPEN OPTOlogic, Grant No 899794, EU Horizon Europe Program (grant agreement No 101080086 NeQSTGrant Agreement 101080086 — NeQST); ICFO Internal “QuantumGaudi” project; European Union’s Horizon 2020 program under the Marie Sklodowska-Curie grant agreement No 847648;
“La Caixa” Junior Leaders fellowships, La Caixa” Foundation (ID 100010434): CF/BQ/PR23/11980043.
The work of J.Z. was funded by the National Science Centre, Poland, project 2021/03/Y/ST2/00186 within the QuantERA II Programme that has received funding from the European Union Horizon 2020 research and innovation programme under Grant agreement No 101017733.  A support by the Strategic Programme Excellence Initiative (DIGIWorkd) at Jagiellonian University is acknowledged.
We gratefully acknowledge Polish high-performance computing infrastructure PLGrid (HPC Center: ACK Cyfronet AGH) for providing computer facilities and support within computational grant no. PLG/2024/017289.
The iDMRG simulations have been performed using the TeNPy library \cite{tenpy}.
Views and opinions expressed in this work are, however, those of the authors only and do not necessarily reflect those of the European Union, European Commission, European Climate, Infrastructure and Environment Executive Agency (CINEA), or any other granting authority. Neither the European Union nor any granting authority can be held responsible for them.

\bibliography{GaugedHamiltonian_v3.bbl}

\end{document}